# Lattice vibrations boost demagnetization entropy in shape memory alloy

P.J. Stonaha[1*], M.E. Manley[1], N.M. Bruno[2], I. Karaman[3], R. Arroyave[3], N. Singh[4], D.L. Abernathy[5], S. Chi[5]

[1]*Materials Sciences & Technology Division, Oak Ridge National Lab, Oak Ridge, TN 37831*
[2]*Department of Mechanical Engineering, Texas A&M University, College Station, TX 77843*
[3]*Department of Materials Science and Engineering, Texas A&M University, College Station, TX 77843*
[4]*Department of Engineering Technology, University of Houston, Houston, TX 77004*
[5]*Quantum Condensed Matter Division, Oak Ridge National Lab, Oak Ridge, TN 37831*

## Abstract

Magnetocaloric (MC) materials present an avenue for chemical-free, solid state refrigeration through cooling via adiabatic demagnetization. We have used inelastic neutron scattering to measure the lattice dynamics in the MC material $Ni_{45}Co_5Mn_{36.6}In_{13.4}$. Upon heating across the Curie Temperature ($T_C$), the material exhibits an anomalous increase in phonon entropy of $0.22 \pm 0.04\ k_B$/atom, which is ten times larger than expected from conventional thermal expansion. This transition is accompanied by an abrupt softening of the transverse optic phonon. We present first-principle calculations showing a strong coupling between lattice distortions and magnetic excitations.

The coupling between magnetism and lattice excitations has been a recent topic of interest due to its importance in several solid-state systems, such as influencing decoherence in quantum magnets [1] and inducing a spin-Peierls transition in $CuGeO_3$ [2, 3] and $ZnCr_2O_4$ [4]. Studies of magnetoelastic materials have focused on metamagnetic shape memory alloys (MMSMA), which are able to reversibly deform with strains of up to ~6% [5, 6] via a structural first-order phase transition (FOPT) [7-9]. The FOPT allows the alloy to function as magnetocaloric (MC) material, in which an adiabatic change in magnetization of the MMSMA causes an increase in its entropy, thus lowering its temperature [9-12]. While the change in entropy can be quite large, the FOPT is often accompanied by thermal and/or magnetic hysteresis, which limits the material's long term cycling efficiency. In addition, materials requiring a magnetic field of more than 2 T to induce the FOPT are not feasible for use in any magnetic refrigerator using permanent magnets [13, 14].

The paramagnetic-to-ferromagnetic second-order phase transition (SOPT) offers a means to extract heat from a magnetic material using weaker applied magnetic fields and without hysteresis. This SOPT is present in *any* magnetic material at its Curie temperature ($T_C$), but the change in entropy is typically significantly less than across the FOPT.

*Corresponding author email: stonahapj@ornl.gov

In this letter, we quantify the phonon entropy and the magnon-phonon coupling in $Ni_{45}Co_5Mn_{50-x}In_x$ ($x$=13.4) near the austenite $T_c$ using neutron scattering [15]. When annealed at temperatures below 900K, this compound forms a Huesler L2$_1$ structure that is stable down to the martensitic transition temperature $T_M$, with Mn atoms occupying the 4a positions, $Mn_{1-x/25}In_{x/25}$ the 4b, and $Ni_{0.9}Co_{0.1}$ the 8c positions (Wyckoff notation) [16]. The composition of the crystal puts it at a morphotropic phase boundary; decreasing $x$ from 13.4 to 13.3 leads to an increase in $T_M$ by ~100K [17]. *Ab initio* calculations of similar compounds point to magnetic frustration as a possible explanation of the morphotropic phase boundary [18-21]. The magnetic exchange integral $J_{ij}$ for nearest-neighbors Mn-Mn in the 4a and 4b positions – i.e. aligned along [100] – is predicted to be a large negative value (antiferromagnetic). Such an interaction competes with the predicted positive $J_{ij}$ [22] between nearest-neighbor Co-Mn and Ni-Mn atoms aligned along [111].

We grew single crystals of $Ni_{45}Co_5Mn_{36.6}In_{13.4}$ using the Bridgeman technique, which we then homogenized under a protective argon atmosphere at 1173 K for 24 hours. The alloys were furnace (slowly) cooled across the ordering temperature (900 K) to promote L2$_1$ crystallographic ordering. The finished sample comprised 6 co-aligned single crystals arranged in a thin plane measuring 15 mm x 25 mm x 0.5 mm. The thin-plate geometry was chosen to minimize neutron absorption from the In atoms for most scattering angles, while maximizing the amount of material in the neutron beam during the inelastic neutron scattering measurements of the phonon dispersion curves. To obtain the phonon density of states, an additional 1000 mm$^3$ of the crystal was pulverized to a ~75 μm grain size, re-annealed at 1173K for 2 hours to relieve internal lattice strain, and then furnace cooled. SQUID magnetometry measurements indicate that $T_C$ and the $T_M$ of both the crystals and the powder are 395 K and 230 K, respectively (see Supplemental Material, S.1).

The phonon densities of states (DOS), extracted from measurements of the $Ni_{45}Co_5Mn_{36.6}In_{13.4}$ powder, are presented in Fig. 1(a). We made these measurements with 50 meV incident energy neutrons using the ARCS time-of-flight neutron spectrometer at the Oak Ridge National Lab (ORNL) at temperatures of 300 K and 450 K. Scattering from the sample enclosure was subtracted, and the data were corrected for multiphonon contributions using an iterative procedure [23]. The DOS shows a large shift towards low energy at the higher temperature, especially at the peak near 10 meV. The change in entropy was calculated using a force constant model described *below*.

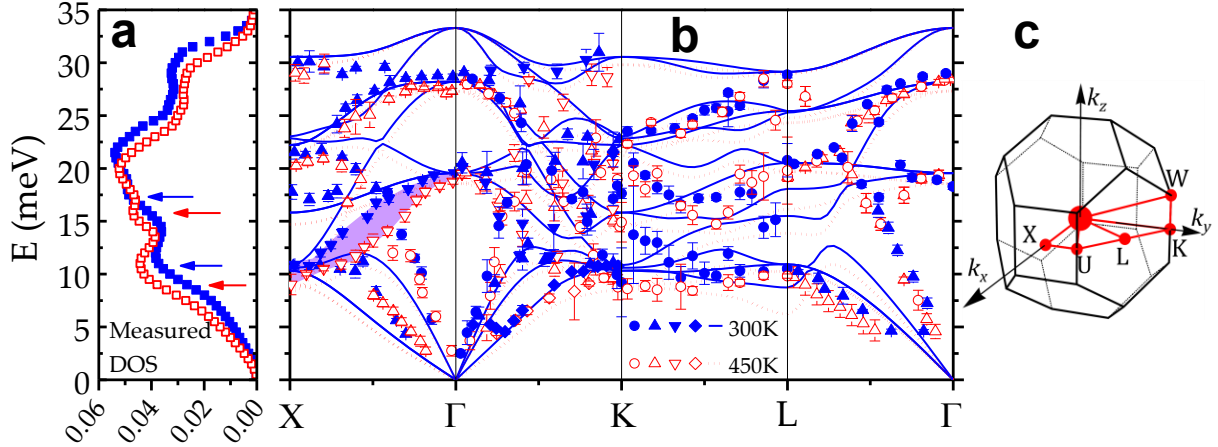

FIG. 1. Phonons in Ni$_{45}$Co$_5$Mn$_{50-x}$In$_x$ ($X = 13.4$): (a) Measured DOS from the powder sample at 300K (solid blue) and 450K (open red). The arrows indicate the boundaries of the largest phonon softening. The point size represents the largest uncertainty; (b) Phonon dispersion measurements (points) and model calculations (lines) from the single crystal sample at 300K (solid blue) and 450K (open red). The shaded region along X → Γ represents the region of largest phonon softening. The phonon polarizations are indicated by the point shapes: longitudinal (circle), transverse along even/odd Miller indices (triangle/upside-down triangle), and [1$\bar{1}$0]-polarized (diamond). Data along K→L is limited due to indium absorption self-shielding. Error bars represent the quadrature sum of the counting statistics' uncertainty and the standard deviation of rebinned data points.; (c) the Brillouin zone, representing the scattering geometry on ARCS.

Inelastic neutron scattering from the single crystals was measured on the ARCS spectrometer at the Spallation Neutron Source at ORNL. Measurements were made with the sample 300 K and 450 K (see S.4 for experiment details). The geometry of the ARCS instrument allowed detection of scattered neutrons out to 5$^{th}$-order Bragg peaks in the {1$\bar{1}$0} plane and within the first zone in the [1$\bar{1}$0] direction. The location of the Bragg peaks in $\boldsymbol{Q}$-space remained constant throughout the experiment, indicating that the crystal did not undergo a structural phase change. From the elastic scattering, we calculated the lattice parameter to be $a_0 = 5.951 \pm 0.003$ Å at both temperatures. We determined the phonon dispersions using a fitting routine described in S.3. The results of this process are shown in Fig. 2(b).

We modeled the phonons using the PHONON software package [24]. The modeled unit cell was a full Heusler with L2$_1$ symmetry (space group $Fm$-$3m$) comprising three atom types – Mn, In, and Ni*, where Ni* atoms have a mass equal to the composition-weighted average mass of Co and Ni atoms. The lattice parameter was set to $a_0 = 5.95$Å. The model calculations were done using a shell model with one force parameter to describe each atom-atom interaction. Achieving good agreement between the modeled and measured DOS required 9 force parameters. The force constants used to match the data taken at $T = 300$ K were independent from the force constants used for the $T = 450$ K data (see S.2). Agreement between the modeled and measured phonon dispersions is generally good, although the model predicts softer acoustic phonons than were measured.

We extracted thermodynamic quantities using the neutron scattering data. The phonon specific heat $C(T)$ was computed from each DOS $g(E)$ (c.f. Fig. 1(b)) as

$$C(T) = 3k_B \times \int_0^{E_c} g(E) \left(\frac{E}{k_B T}\right)^2 \frac{\exp\left[\frac{E}{k_B T}\right]}{\left(\exp\left[\frac{E}{k_B T}\right]-1\right)^2} dE, \quad (1)$$

where $E_c$ is the cut-off energy. The change in vibrational entropy $\Delta S_{vib}$ going from the low-temperature ferromagnetic (FM) state to the high-temperature paramagnetic (PM) state was then estimated by integrating

$$\Delta S_{vib} = \int_0^{T_c} \left(\frac{C_{PM}(T)}{T} - \frac{C_{FM}(T)}{T}\right) dT. \quad (2)$$

Using the data, we calculate $\Delta S_{\text{vib}} = 0.28\ k_B(\text{atom})^{-1}$. After accounting for the element-specific mass and neutron scattering cross-section (c.f. S.3), we conclude a true change in vibrational entropy of $0.24 \pm .04\ k_B(\text{atom})^{-1} \approx 31 \pm 5\ \text{J}\ (\text{kg K})^{-1}$. The uncertainty in the change in entropy is determined by varying the force parameters over a range in which the fit to the phonon dispersion remains reasonable.

An equal change in vibrational entropy due to thermal expansion alone would require – assuming a bulk modulus of $B = 100$ GPa (for NiTi, $B = 114$ GPa [25]) – a linear dilation of the unit cell of 1.6%. For a typical metal with a linear thermal expansion coefficient of $\alpha \approx 10^{-5}\ K^{-1}$, such a dilation would require an increase in temperature of ~1600 K. Therefore, we estimate that only about 10% of $\Delta S_{\text{vib}}$ is from normal thermal expansion (over 150 K) and entropy of $0.22\ k_B\ (\text{atom})^{-1} \approx 28\ \text{J}\ (\text{kg K})^{-1}$ remains unaccounted for. There is no structural phase change but rather a magnetic transition over the measured temperature range, suggesting that the unaccounted-for change in vibrational entropy arises from magnetostrictive effects and/or magneto-phonon coupling. The $0.22\ k_B(\text{atom})^{-1}$ of anomalous entropy change is ~20% greater than the reversible entropy change that has been measured across the martensitic FOPT in similar MMSMAs [17, 26-29], presenting this crystal as a possible candidate for a SOFT-based magnetocaloric material.

To explore more directly the relationship between the lattice vibrations and the magnetic structure, we performed temperature-dependent triple-axis inelastic neutron scattering measurements of key phonons across the magnetic phase transition. From the ARCS data, there appears a large temperature dependence of the energy of the lowest-energy TO [ξ00] phonon near the zone edge (indicated by the arrows in Fig. 1(a) and shaded region in Fig. 1(b)). Using the same crystal from the ARCS experiment, we measured the [ξ00] phonon dispersion near [330]. These measurements were made on the HB3 instrument at the High Flux Isotope Reactor at Oak Ridge National Lab (ORNL) with the temperature scanned from 230 K – 500 K using a cryodisplex (see S.5).

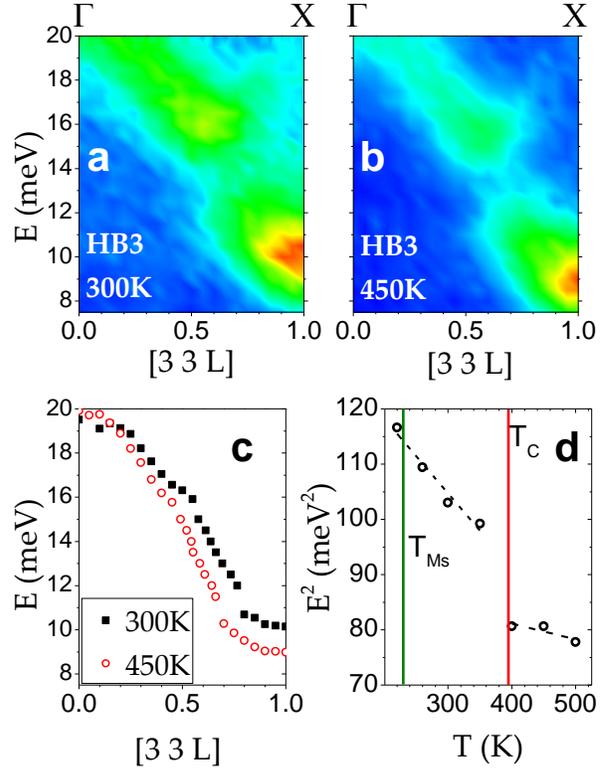

FIG. 2. Triple-axis inelastic neutron scattering data from $Ni_{45}Co_5Mn_{50-x}In_x$ ($X = 13.4$): (a,b) the color plots of the raw scattering data at temperatures above and below $T_c$ ($T_c = 395K$); (c) the energies of the peak in the phonon; (d) The energy of the phonon at the X-point vs. temperature. Dashed lines are best fits, mentioned in text. The Curie and martensitic transformation temperatures are indicated with vertical lines.

The TO phonon dispersions at temperatures of $T = 300$ K and 450 K are presented in Fig. 2(a,b), respectively. For both temperatures, the energy of the TO phonon is 20 meV at the Γ-point and decreases with increasing $q$. We extracted one-dimensional cuts in constant $q$ and E in the same manner as the data from ARCS (c.f. Fig 2(c)). The value of the phonon energy at the zone edge (X-point) is strongly temperature-dependent (c.f. Fig. 2(d)), falling from 10.2 meV at 300 K to 9.0 meV at 450 K. This represents a fraction change in energy of ~12%. The softening from thermal expansion can be estimated as $\Delta\omega/\omega = -3\gamma\alpha\Delta T$, where $\Delta T$ is the change in temperature and $\gamma$ is the Grüneisen parameter. The value of $\gamma$ is commonly about 2 [30, 31], and thus the expected decrease in phonon energy from normal thermal expansion is only ~0.9%; as with the DOS, the change in energy of this TO phonon mode cannot be attributed to thermal expansion alone. At the zone edge, the temperature-dependence of the phonon frequency scales linearly in energy squared (c.f. Fig. 2(d)). Above and below $T_c$, the energy follows $-0.03 \pm 0.02$ meV$^2$/K and $-0.14 \pm 0.02$ meV$^2$/K, respectively, with a jump of 20 meV$^2$ across a narrow temperature range near $T_c$ (this represents an absolute change of 0.75 meV). The abrupt change in the zone-edge energy upon crossing $T_c$ and the steeper slope in the ferromagnetic region confirm the conclusion from the above thermodynamic argument that strong magnetoelastic coupling is occurring in the crystal.

The shape of the low energy TO [$\xi$00] phonon in Fig. 2(a,b), in particular the break in intensity near $E = 13$ meV, appears similar to an avoided crossing of local-mode vibrations, which is observed in the excitation spectrum of other inhomogeneous materials [32-36].

We recognize that at the zone edge, the TO [$\xi$00] mode is identical to the LA[$\xi\xi$0] mode; any softening observed in the former must also exist in the latter. In the ARCS single crystal scattering data, we do indeed observe a softening of the LA[$\xi\xi$0] mode, but this softening is limited to an area of reciprocal space very close to the zone edge (see S.5). The magneto-phonon coupling is apparently strongest for oscillations in the transverse direction.

To further understand the phonon-magnon coupling, we calculated the magnons in the crystal using the first-principles SPR-KKR method [37, 38]. The model structure was similar to the PHONON model described above, with modifications allowing for Mn atoms to occupy the In sublattice and Co atoms to occupy the Ni sublattice. The structure was assumed to have an antiferromagnetic ground state. For each calculation, we first calculated the self-consistent potential, which we then used to calculate the exchange interactions. From the exchange interactions, magnons were calculated using the procedure reported in Ref. [39].

The [$\xi$00] magnon dispersion is shown in Fig. 3(a) for the L2$_1$ structure with lattice parameters between $a_0 = 5.91$Å and $a_0 = 6.00$Å. Three stable magnons appear from the calculation – 2 optic and 1 acoustic. The frequency of the highest energy optic magnon is highly sensitive to the lattice parameter. The unstable magnon is the result of the strong Mn-

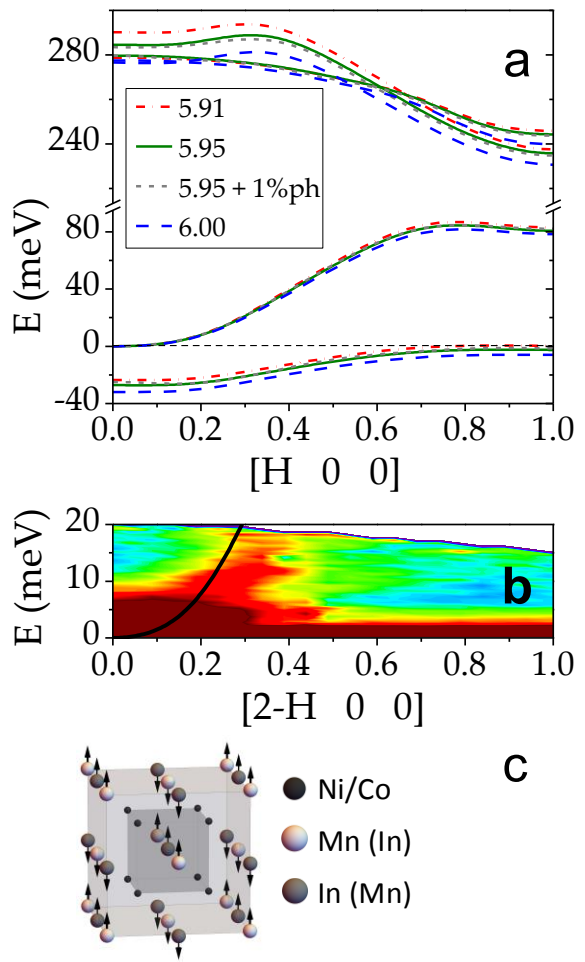

FIG. 3. First principles magnon calculations: (a) magnon dispersion for lattice parameters $a_0 = 5.91$Å (red, dash-dot), $5.95$Å (green, solid), $6.00$Å (blue, dash), and $5.95$Å with a 1% frozen phonon distortion (black, dotted); (b) image plot of the measured inelastic neutron scattering along the [$\xi$00] direction with the sample at 300K. The overlaid black lines corresponds to the calculated magnon for $a_0 = 5.95$Å.; (c) cartoon of the supercell, with arrows indicating displacement directions of the 1% frozen phonon distortion used in (a). Atoms on the Ni/Co, Mn, and In sublattices are shown in black, white, and grey, respectively.

Mn antiferromagnetic interactions arising from RKKY and superexchange interactions [22] and may be a signature of the instability of collinear magnetic states against more complex non-collinear configurations [40]. The calculated acoustic [$\xi$00] magnon dispersion matches well with the observed magnon in the region $0 < q < 0.4$ (c.f. Fig. 3(b)).

Given that the largest observed phonon softening is focused at a zone edge, the change in the force constants is likely the result of antiferromagnetic interactions. We hypothesize that the stiffening of the TO[$\xi 00$] below $T_C$ is due to magnetic interactions between Mn-Mn atoms displaced in the [$\xi 00$] direction. To test this hypothesis, we performed a first-principles "frozen phonon" calculation of the magnon dispersion with frozen displacements of the Mn/In atoms along the zone-edge eigenmode shown in Fig. 3(c). Atoms on the Mn sublattice were displaced by $0.01 \times a_0$, while those on the In sublattice were displaced by $0.01\sqrt{f} \times a_0$, respectively, where $f = M_{Mn}/M_{In} \approx 0.45$ is the ratio of atomic masses of Mn and In. The results of the calculation show this frozen phonon has little effect on the [$\xi 00$] magnon dispersion (c.f. Fig. 3(a)). The interactions between the phonons and magnons may be much more complex than can be encompassed by a single frozen phonon calculation, which would not be surprising given that the phonon softening is observed across the entire reciprocal space, as seen in Fig. 1(b). Correctly calculating the magnetoelastic coupling may require a full *ab initio* molecular dynamics calculation with magnetic interactions.

The vibrational DOS and the phonon dispersion have been measured in $Ni_{45}Co_5Mn_{50-X}In_X$ ($X = 13.4$) using neutron scattering. A shell model calculation of a 3-atom full Heusler crystal was sufficient to reproduce the observed phonon dispersion. From the calculated DOS, we estimate that the excess change in vibrational entropy across $T_C$ due to magnetoelastic coupling is $0.22\ k_B\ (atom)^{-1}$. The measurements of the phonon dispersion indicate that the low-energy TO [$\xi 00$] phonon couples strongly to the magnetic structure in this ferro/paramagnetic crystal, and the first-principles calculations indicate magnetostriction occurring in the material.

This work presents the first quantified measurement of the change in phonon entropy due to demagnetization of an MMSMA at the magnetic SOPT. The crystal studied herein exhibits a large increase in vibrational entropy across this transition, presenting an opportunity for its use in magnetic refrigeration. For magnetic refrigeration, adiabatic demagnetization is controlled with an external field, rather than by crossing $T_C$, and the phonon entropy made available in the process adds to the increase in magnetic entropy.

**Acknowledgements:** Research sponsored by the U.S. Department of Energy, Office of Basic Energy Sciences, Materials Sciences and Engineering Division. The portions of this research performed at the Oak Ridge National Laboratory's Spallation Neutron Source and High Flux Isotope Reactor facilities were sponsored by the U. S. Department of Energy, Office of Basic Energy Sciences. IK, NS, and RA acknowledge support of NSF under grants No. DMR-0844082 and DMR-0805293. First principles calculations were carried out at the TAMU Supercomputing Facility.

# Lattice vibrations boost demagnetization entropy in shape memory alloy

*P.J. Stonaha, M.E. Manley, N.M. Bruno, I. Karaman, R. Arroyave, N. Singh, D.L. Abernathy, S. Chi*

## Supplemental Material

### 1. Magnetometry measurements of $Ni_{45}Co_5Mn_{36.6}In_{13.4}$

SQUID magnetometry was performed on a Quantum Design Magnetic Property Measurement System (MPMS3). Single crystals and powders were heated to 400K under zero magnetic field and subsequently field cooled (FC) and field heated (FH), and the results are shown in Fig S.1. The MMSMA sample temperature was ramped at 5 K/min in a rough vacuum/He gas environment (5 Torr).

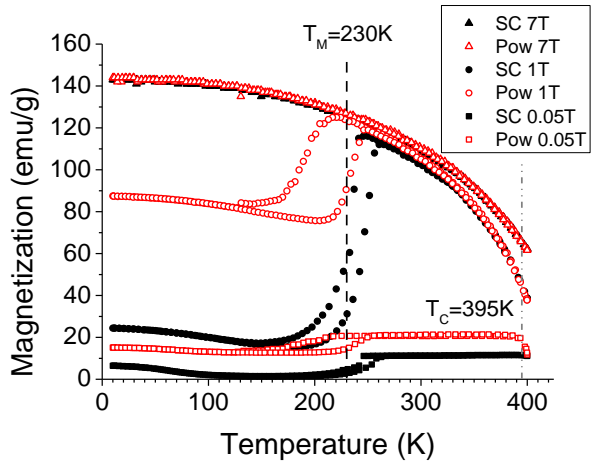

**FIG. S.1:** Thermomagnetization curves of the single crystal (SC - black, filled) and powder (Pow – red, open) samples at applied magnetic fields of 0.05 T, 1 T, and 7 T (squares, circles, and triangles respectively).

### 2. PHONON model

The PHONON model calculation was carried out with a 2x2x2 supercell. The force parameters used are listed in Table S.1. All force parameters beyond eighth-nearest neighbor were explicitly set to zero.

| Atoms | Ni-Mn | Ni-In | Ni-Ni | Mn-In | Ni-Ni | Mn-Mn | In-In | Ni-Mn | Ni-In |
|---|---|---|---|---|---|---|---|---|---|
| Interatomic distance (Å): | 2.5764 | 2.5764 | 2.9750 | 2.9750 | 4.2072 | 4.2072 | 4.2072 | 4.9335 | 4.9335 |
| F.C. @ 300K | 8000 | 10000 | 7000 | 8750 | -2250 | -3750 | 0 | 2000 | -1500 |
| F.C. @ 450K | 8000 | 9000 | 6500 | 8750 | -2500 | -3750 | -500 | 2000 | -1500 |

**TABLE S.1:** List of the force constants (F.C.) used in the PHONON model calculation. The units of the force constants are u THz$^2$.

### 3. Calculation of vibrational entropy

We calculated a T=300K DOS using the same PHONON model that produced the T=300K dispersion curves. The calculated DOS was weighted with a Gaussian resolution function with an energy-dependent width $\Omega(E) = 3.5 * (1 - E/E_i)^{3/2}$. The modeled system contained an equal composition of In and Mn; to more accurately match the studied system, the calculated partial DOSs of In and Mn were scaled in intensity by 0.134/0.25 and 0.366/0.25, respectively. The calculated neutron-weighted DOS was then scaled in energy to reproduce the entropy change observed in the neutron scattering, shown in Fig. S.2(a). The scaling function used was $I_0 + (I_1 - I_0) * \omega/\omega_{Max}$, representing a multiplicative shift of $I_0$ at $\omega = 0$ and $I_1$ at $\omega = \omega_{Max}$. The values of $I_0$, $I_1$, and $\omega_{Max}$ were set to 0.665, 1, and 35 meV, respectively. This same scale function was then applied to the calculated true DOS (c.f. Fig S.2(b-c)). The true change in entropy was calculated as the difference in the shifted and unshifted true DOSs.

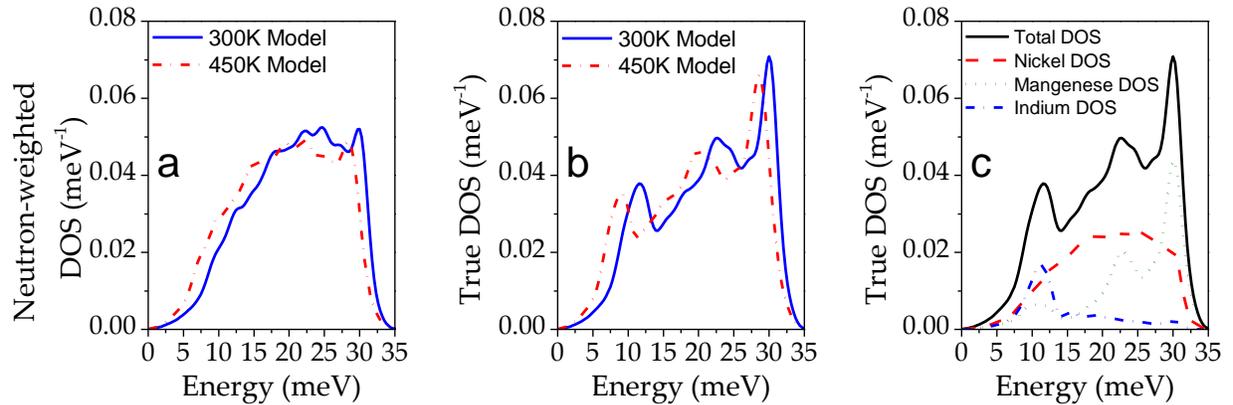

Fig S.2: Calculated densities of states (DOSs) from the PHONON model: (a) neutron-weighted DOS; (b) true DOS. In (a) and (b), the 300K model (blue, solid) is produced from the PHONON software, and the 450K model (red, dash-dot) is produced by scaling the 300K model.; (c) partial true DOSs, showing the contribution from the different elements – Ni (red, dash), Mn (green, dot), In (blue, dash-dot), and total DOS (black, solid).

### 4. Inelastic neutron scattering experiment on ARCS

Measurements on the ARCS spectrometer were made with fixed incident neutron energy of $50 meV$. The crystal was rotated from 40° to 125° about [1$\bar{1}$0], where 0° denotes the incoming neutron beam is aligned with [001]. Temperature values mentioned in text have an uncertainty of $\pm 15\ K$. The ARCS instrument collects 4-dimensional ($\mathbf{Q}, E$) data sets. Two-dimensional ($q$,E) slices in the data were taken along high symmetry directions, and one-dimensional cuts in both constant $q$ and constant E were extracted from the data and individually fit to a sum of Lorentzian curves, a Gaussian curve at the elastic peak, and a linear background. We repeated this process across all zones for which there was sufficient scattering data. The locations of the center of the Lorentzian curves were filtered by their fitted uncertainty and averaged with the locations of peaks at nearby ($q$,E) points across all other zones.

## 5. Inelastic neutron scattering experiment on HB3

The HB3 instrument was operated with a variable vertically focusing PG002 monochromator (d-spacing = 3.355 Å). Collimation from the premonochromator, monochromator-to-sample, sample-to-analyzer, and analyzer-to-detector were set to 48′, 60′, 80′, and 120′ FWHM, respectively. We made measurements along constant $E$ cuts with a fixed initial neutron energy of 14.7 meV. Data was collected in steps of 0.5 meV and 0.05 rlu in $E$ and $q$, respectively. The data presented in this work are from the high-resolution side of the resolution ellipse. The temperature of the sample was scanned in the following order: $300K \rightarrow 260K \rightarrow 350K \rightarrow 400K \rightarrow 450K \rightarrow 500K \rightarrow 220K$. Temperature values have an uncertainty of $\pm 15$ K.

## 6. LA [110] mode

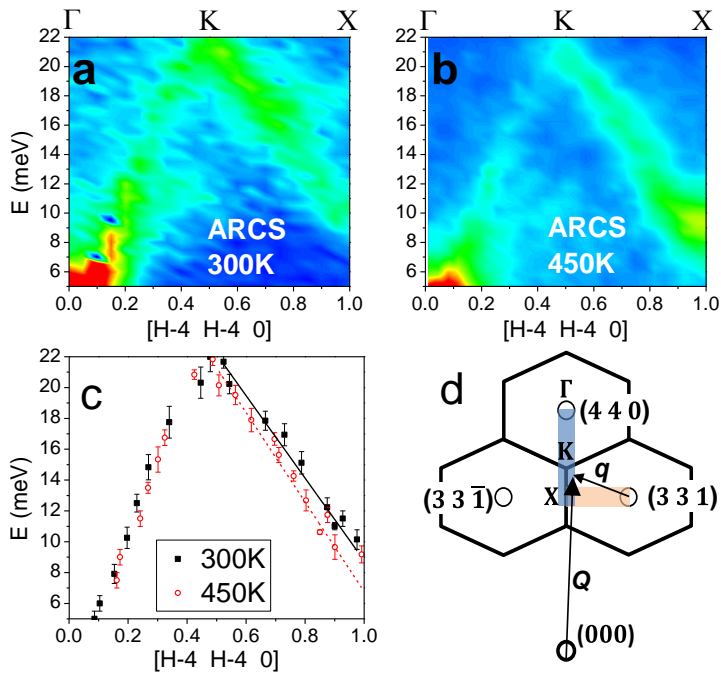

FIG. S.3: Inelastic neutron scattering data of the LA [110] mode measured on ARCS: (a,b) Color plots of the scattered intensity; (c) the locations of the centers of Lorentzian functions fit to the data ; (d) the reciprocal lattice representing the plane of scattering in this figure (blue region) and in Figure 3 (orange region). High symmetry points are indicated.

## 7. Prediction of Magnetic Exchange Constants

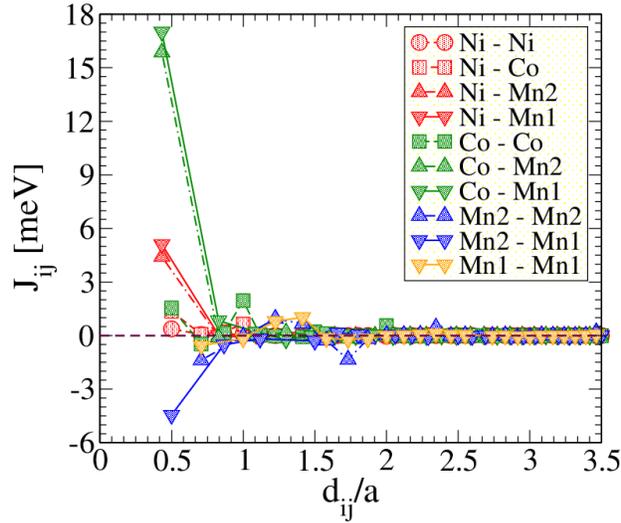

**FIG S.4: Magnetic exchange interaction ($J_{ij}$) of L21 unit cell of $Ni_{45}Co_5Mn_{36.6}Mn_{13.4}$ for lattice parameter, $a = 5.95$Å. Mn1 denotes the Mn atom on the host sub-lattice, Mn2 denotes the Mn atoms on the In sub-lattice. $J_{ij}$'s for the In atom have not been plotted, as their magnitude is negligible.**

Figure S.4 shows the calculated magnetic exchange constants, $J_{ij}$, for the system under study. Due to mainly RRKY-type interactions the interactions between Mn atoms in the Mn- and In-sublattices is antiferromagnetic (FM). Note that while ferromagnetic (FM) Co-Mn interactions are much higher than those corresponding to the Ni-Mn magnetic exchange. However, Co exists at relatively low compositions and its effect on the magnetic structure of the system is expected to be small. When comparing FM Ni-Mn vs AFM Mn-Mn interactions, we can observe that they are almost of equal magnitude. This competition may be responsible for the stabilization of non-collinear magnetic states at low temperatures, as suggested by the calculated magnon dispersions shown in Fig. 3. Note that there is experimental evidence---as mentioned in [22]---that alloys of this composition exhibit magnetic glassy behavior at very low temperatures and these states are by necessity non-collinear.